\documentclass[aps,prl,showpacs,twocolumn,superscriptaddress]{revtex4}
\usepackage{mathrsfs}
\usepackage{epsfig}
\usepackage{graphicx}
\usepackage{amsfonts}
\usepackage[figuresright]{rotating}
\usepackage{amssymb}
\usepackage{amsmath}
\usepackage{dcolumn}
\usepackage{bm}
\usepackage{color}

\def\be{\begin{equation}} \def\ee{\end{equation}}
\def\bea{\begin{eqnarray}} \def\eea{\end{eqnarray}}

\newcommand{\lmu} {Department of Physics and Arnold Sommerfeld Center for Theoretical Physics,
Ludwig-Maximilians-Universit{\"a}t M{\"u}nchen, Theresienstr.\ 37,
80333 Munich, Germany}

\newcommand{\iqoqi} {Institute for Quantum Optics and Quantum Information, Austrian Academy of Sciences, 6020 Innsbruck, Austria}

\begin{document}
\title{Identifying a bath-induced Bose liquid in interacting spin-boson models}

\author{Zi Cai}
\affiliation{\lmu}
\affiliation{\iqoqi}

\author{Ulrich Schollw\"{o}ck}
\affiliation{\lmu}

\author{Lode Pollet}
\affiliation{\lmu}

\begin{abstract}
We study the ground state phase diagram of a one-dimensional hard-core bosonic model with nearest-neighbor interactions (XXZ model) where every site is coupled Ohmically to an independent but identical reservoir, hereby
generalizing spin-boson models to interacting spin-boson systems.
We show that a  bath-induced Bose liquid phase can occur in the ground state phase diagram away from half filling. This phase is compressible, gapless, and conducting but not superfluid.
At haf-filling, only a Luttinger liquid and a charge density wave are found. The phase transition between them is of Kosterlitz-Thouless type where the Luttinger parameter takes a non-universal value.
The applied quantum Monte Carlo method can be used for all open bosonic and unfrustrated spin systems, regardless of their dimension, filling factor and spectrum of the dissipation as long as the quantum system couples to the bath via the density operators.

\end{abstract}


\maketitle

{\it Introduction --} Quantum systems are, in general,
coupled with their surroundings.   In standard textbook
scenarios it is assumed, explicitly or implicitly, that
the system-environment coupling is weak such that the
equilibrium state of the system can be described by the
Boltzmann-Gibbs ensemble. However, this property no
longer holds for quantum systems with finite coupling strength to the environment  (i.e., when this coupling is comparable with the typical energy scales in the system), where the system-environment coupling can
qualitatively change the properties of the system\cite{Hanggi2008}. The paradigmatic
model of quantum open systems is the spin-boson model \cite{Caldeira1981, Caldeira1983,Caldeira1983b},: a
two-level (spin-1/2) system coupled to a bath of harmonic
oscillators with an infinite number of bosonic degrees of
freedom\cite{Leggett1987,Grabert1988,Weiss1999}. The coupling to
the heat bath drives a transition between a localized (classical) and delocalized (quantum) state
for the spin, which is closely related with
the long-range Ising model\cite{Dyson1969,Kosterlitz1976} and
quantum impurity models\cite{Anderson1970,Emery1974,Chakravarty1982,Bray1982,Guinea1985}.

Though systems consisting of a single or a few spins coupled to a heat bath have been
discussed extensively\cite{Leggett1987,Weiss1999,Vojta2006,Hur2008,Bulla2008},
the situation is much more complicated when the
system itself is an interacting quantum many-body system.
The interplay between many-body effects and dissipation
opens avenues for observing unknown and richer phenomena~\cite{Chakravarty1986,Fisher1987,Kapitulnik2001,Dalidovich2002,Neto1997,Cazalilla2006,Diehl2008,Prosen2008,Verstraete2009,Daley2009,Torre2010,Poletti2012,Lobos2011,Lobos2012,Barreiro2010,Barreiro2011,Chen2014,Rancon2013,Schachenmayer2014,Schindler2012,Sieberer2013,Cai2013,Winter2014}
than what is expected on the basis of these effects separately.
Notwithstanding the intrinsic difficulties with strong correlations,
significant progress has been made for fermionic systems with retarded interactions by using determinant QMC methods~\cite{Assaad2007,Hohenadler2012,Raczkowski2010} and dynamical mean field theory~\cite{Werner2010}, and for one-dimensional (1D) open quantum many-body systems using bosonization~\cite{Neto1997,Cazalilla2006,Malatsetxebarria2013}.
Also some specific models such as Ising-like Hamiltonians with site coupling of Ising spins to the bath, or XY-like Hamiltonian with coupling of the type $\sigma^{+}b + \sigma^{-}b^{\dagger}$ have been studied but simulations were typically performed for classical systems~\cite{Werner2004,Werner2005a,Werner2005b,Sperstad2011,Stiansen2012}. However, studies of general (bosonic) quantum models with the density operator coupling to the bath have not been systematically undertaken quantitatively. Quantum Monte Carlo simulations  (QMC) along the lines outlined in this work can in general be applied to such models as long as the system has a positive representation.

In this Letter we apply a  numerically exact QMC method with worm-type updates~\cite{Prokofev1998}  implemented in Ref.~\cite{Pollet2005} (for a recent review, see \cite{Pollet2012_review}) to study the equilibrium properties of
open quantum many-body systems.  Our work is a natural
generalization of previous seminal work on spin-boson models~\cite{Leggett1987,Winter2009} to the many-spin cases,
where each spin not only interacts with a local environment but also with other spins.
Away from half filling we find a gapless, compressible, conducting but non-superfluid phase, which has all the properties of a Bose liquid. Since it only exists thanks to the harmonic bath,  we term it a bath-induced Bose liquid (BIBL).
Throughout this paper we will use the language of hard-core bosons instead of the equivalent spin-$1/2$ terminology.

{\it Model and method --}
We study a 1D lattice of $L$ sites on which hard-core bosons live with system Hamiltonian
\begin{equation}
H_s=\sum_{\langle ij\rangle}\{-t(a_i^\dag a_j+a_j^\dagger a_i)+V (n_i-\frac 12) (n_j-\frac 12) \}-\mu \sum_i n_i, \label{eq:hardcore}
\end{equation}
where $t$ denotes the hopping amplitude, $V$ the nearest-neighbor (NN) density-density interaction strength and $\mu$ the chemical potential (half filling corresponds to $\mu = 0$).
This Hamiltonian is equivalent to the XXZ model with a magnetic field. Our unit is $t=1$. We are interested in the ground state and the critical properties of the quantum phase transitions,  which we will find from a finite size scaling assuming dynamic exponent $z=1$ or $2$ depending on the filling factor.
On each site $i$ the density operator $n^i$ additionally couples to  a local bath  (as in a spin-boson model) resulting in the full Hamiltonian for the system+environment,
\begin{equation}
H=H_s+\sum_{i,k} [\lambda_{ik} (n_i - \frac{1}{2}) (b_{ik}+b_{ik}^\dag)+ \omega_{ik} b_{ik}^\dag b_{ik}], \label{eq:Ham}
\end{equation}
where $b_{i,k}$ and $b_{i,k}^{\dagger}$ denote the annihilation and creation operators of the bath with eigenmodes $\omega_k$ on site $i$ and characterized by the spectral density
\begin{equation}
J(\omega)=\pi\sum_k \lambda_k^2 \delta(\omega-\omega_k)=\pi\alpha
\omega^s  \, \, \, \, ( 0<\omega< \omega_D), \label{eq:ohmic}
\end{equation}
where $\alpha$ represents the coupling strength. The spectral function  $J(\omega)$ is chosen to be linear in $\omega$ corresponding to Ohmic coupling ($s=1$) and has a hard frequency cutoff $\omega_D$ ($\omega_D=10$ in this work), $J(\omega) = 0$ for $\omega > \omega_D$.

The oscillator degrees of freedom can be integrated out yielding a retarded density-density interaction term in imaginary time.
The partition function takes the form
\begin{equation}
Z={\rm Tr} e^{-\beta H} =Z_B\int
\mathscr{D}a^{\dagger}_i \mathscr{D} a_i e^{-\beta H_s-S_{ret}}, \label{eq:partition}
\end{equation}
where $H_s(a, a^{\dagger})$ is the system Hamiltonian and
$Z_B=\texttt{Tr}_{\{ b_{ik}\}}e^{-\beta \sum_{ik} \omega_{ik} b^\dag_{ik} b_{ik}}$ the partition function for the free bosons of the bath. $S_{\rm ret}$ describes
the effective action of the onsite retarded interaction,
\begin{equation}
S_{\rm ret} =-\int_0^\beta d\tau \int_0^\beta d\tau' \sum_i
(n_i (\tau) - \frac{1}{2}) D(\tau-\tau') (n_i(\tau') - \frac{1}{2}), \label{eq:ret}
\end{equation}
with site-independent kernel~\cite{Winter2009}
\begin{equation}
D(\tau-\tau')=\int_0^\infty d\omega \frac{J(\omega)}{\pi}
\frac{\cosh(\frac{\omega\beta}2-\omega|\tau-\tau'|)}{\sinh(\frac{\beta\omega}2)}.
\end{equation}
The asymptotic behavior of the kernel at zero temperature for $\tau\gg \tau_c$
is $D(\tau) \propto 1/\tau^{1+s}$, where
$\tau_c=2\pi/\omega_D$ is the cutoff. For $s \le 1$ and thus including Ohmic dissipation ($s=1$), power counting shows that the retardation is strong enough to induce a transition (cf. the Ising model with long-range interactions $J(x) \sim 1/x^{1+s}$ ~\cite{Dyson1969}). Without dissipation ($\alpha=0$ in Eq.~(\ref{eq:ohmic})),
the XXZ model is free of the sign
problem. Monte Carlo simulations in the presence of dissipation remain possible when keeping the retardation in the exponent. The only change to the implementation of the worm algorithm~\cite{Prokofev1998, Pollet2005} is that the potential energy needs to include the retardation; {\it i.e.}, when the worm is moving around in imaginary time, the evaluation of the integrals resulting from the retardation is required.

{\it Strong dissipative limit -- }
Before analyzing the numerical results, we perturbatively analyze the limit of strong dissipation.
For simplicity we take an XY model ($V=0$). In the limit $t/\alpha\rightarrow 0$ quantum fluctuations are completely suppressed. The system is then in a  mixed state with an equal-weight
mixture of all possible Fock states of hard-core bosons. Half filling requires a more careful analysis beyond this zeroth order result.
Turning on the tunneling but staying  in the regime $t/\alpha\ll 1$, we can treat the tunneling terms as a perturbation, which we restrict to 2nd order virtual hopping processes.
In the dual picture, the world line configuration for the hard-core bosons can be considered as a Coulomb gas
of kinks and antikinks with interactions that are local in space but
long-range in imaginary time, $\mathcal{V}(\tau_i^1-\tau_j^2)\approx -4\alpha \delta_{ij}\ln(|\tau_i^1-\tau_j^2|/\tau_c) $ (that is a 2D Coulombic interaction for a kink-antikink pair located at $\tau_i^1$ and $\tau_j^2$).
The ground-state (kinetic) energy (per site) is then (see the Suppl. Mat.~\cite{Supplementary})
\begin{equation}
E_g/L \approx -\frac{t^2 \tau_c}{4\alpha-1},
\end{equation}
which agrees well with the numerical results as is shown in the Suppl. Mat.~\cite{Supplementary}).
The system can find a lower energy if it can maximize the number of bonds.
Therefore, at half filling, this will require an empty site to be next to an occupied site, since two adjacent empty or  occupied sites can't have virtual exchanges. We expect thus a tendency towards a charge density wave with a gap $\Delta \sim E_g/L$.

\begin{figure*}[htb]
\includegraphics[width=0.3\linewidth]{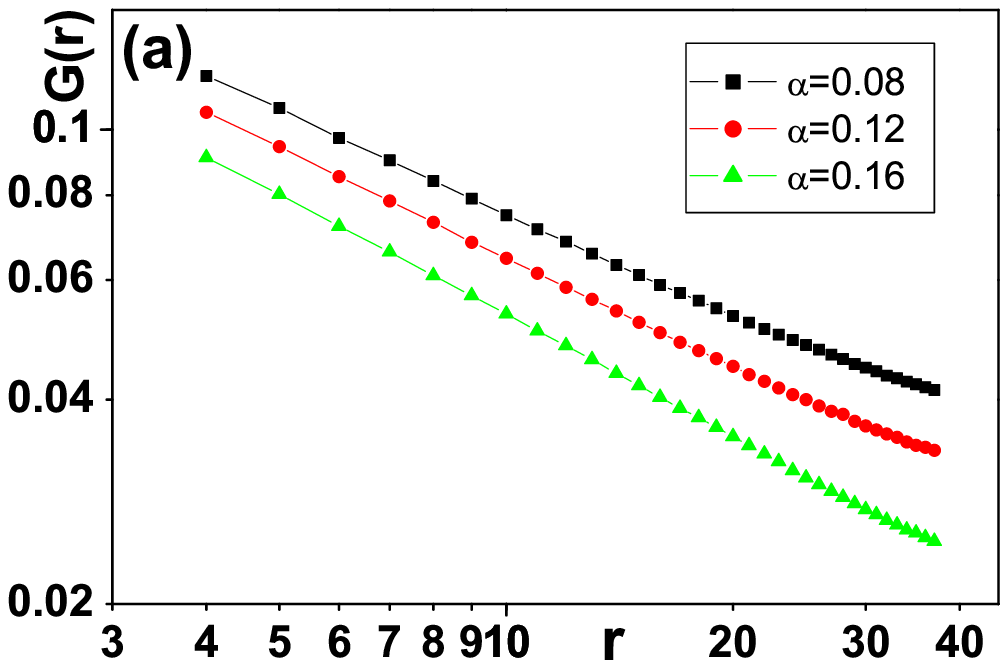}
\includegraphics[width=0.3\linewidth]{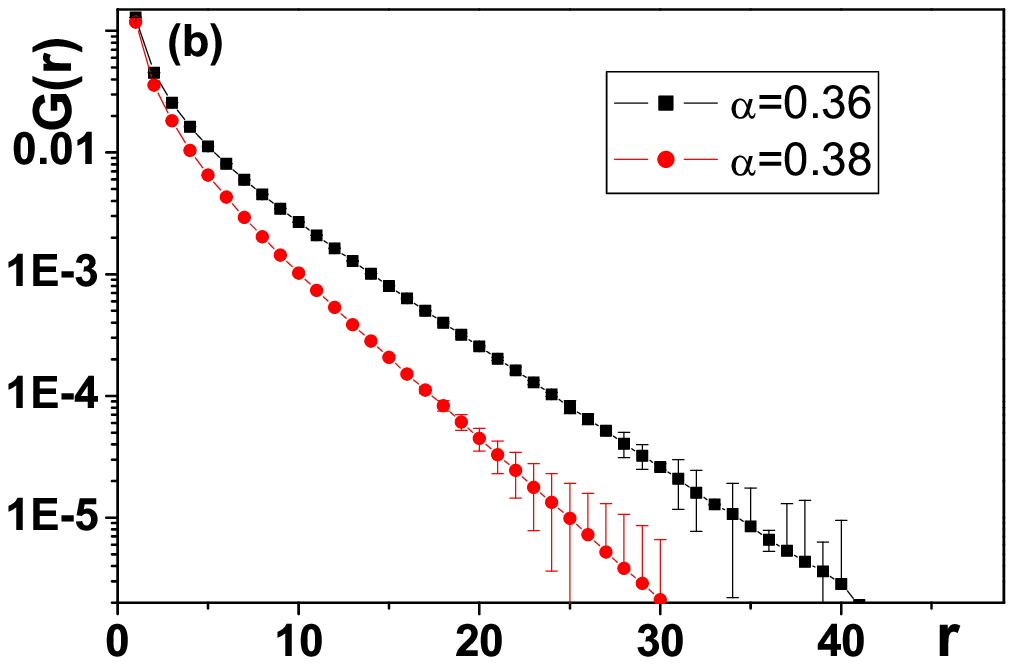}
\includegraphics[width=0.3\linewidth]{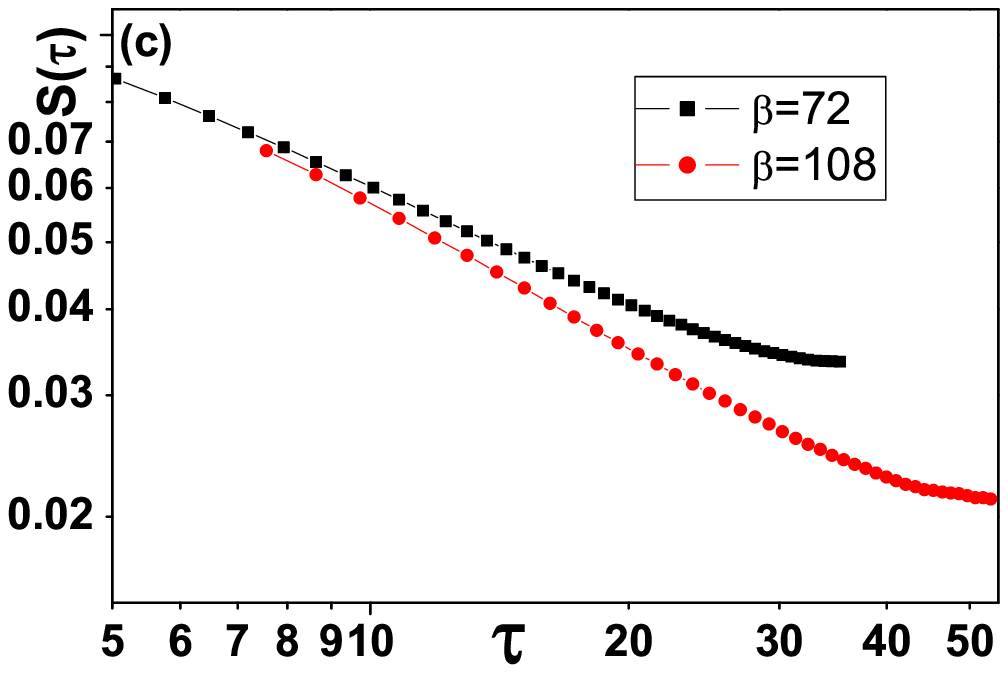}
\includegraphics[width=0.3\linewidth]{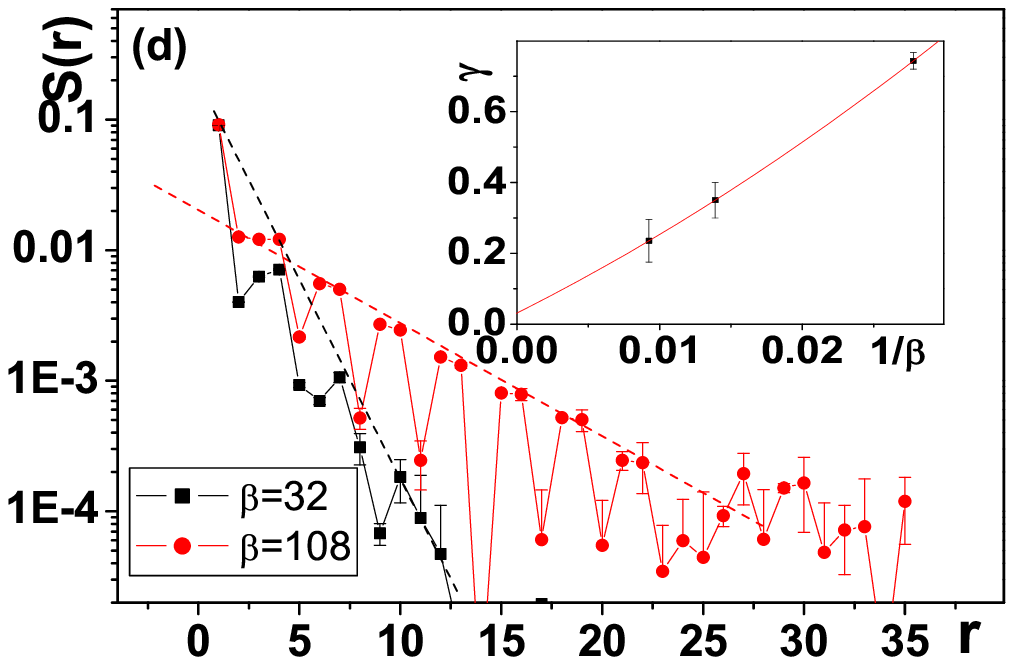}
\includegraphics[width=0.3\linewidth]{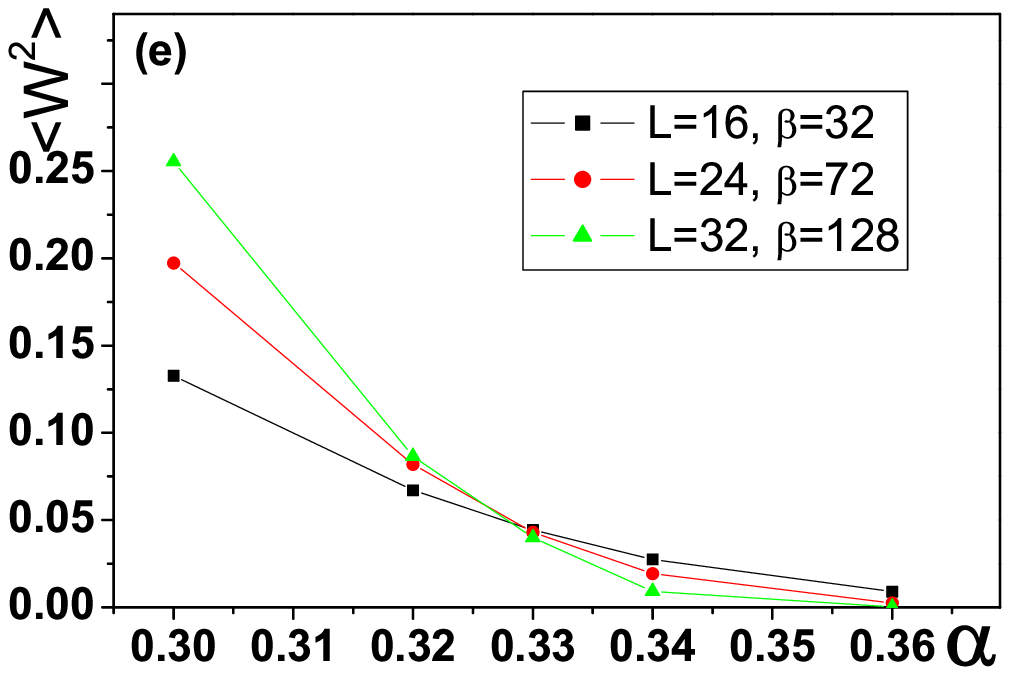}
\includegraphics[width=0.3\linewidth]{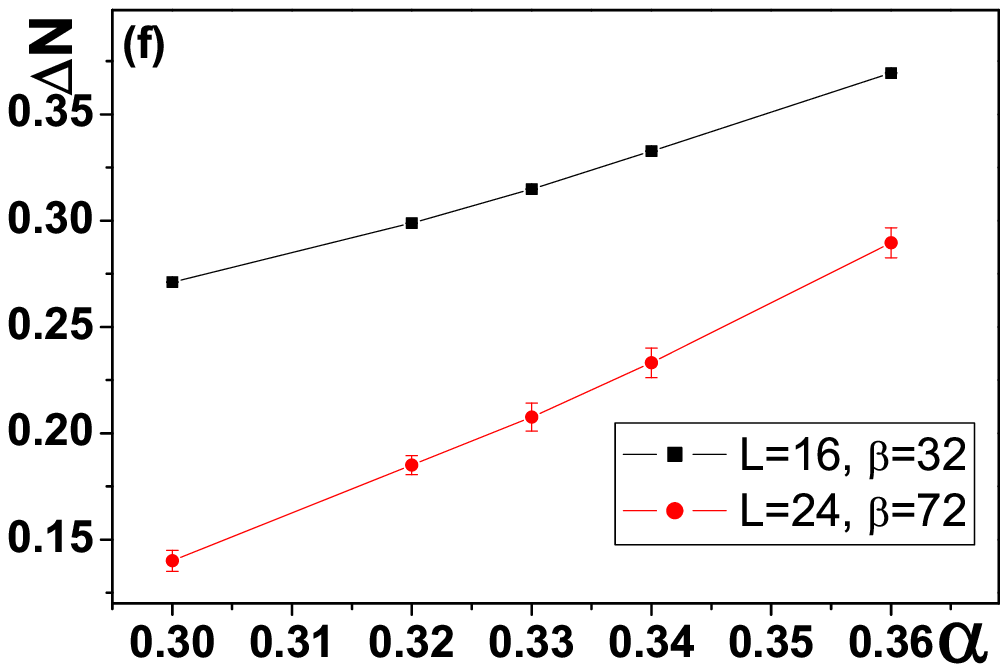}
\caption{(Color online). Single particle correlation function $G(r)$ in (a) the LL phase (algebraic decay); (b) the BIBL phase (exponential
decay) ($L=96$, $\beta=48$, $\mu=-0.1$ for (a)-(b)); (c) unequal-(imaginary) time density correlation functions $S(\tau)$ and (d) equal-time density correlation function $S(r)$ in the BIBL phase for different $\beta$, the inset is the finite-$\beta$ scaling of the exponent $\gamma$;   ($L=72$, $\mu=-0.22$, $\alpha=0.36$ for (c) and (d)); (e) The variance of winding number $\langle W^2\rangle $ and (f) the particle number $\Delta N=\langle N^2\rangle-\langle N\rangle^2 $ as a function of  $\alpha$ with the $z=2$ scaling; ($\mu=-0.1$, $\beta=L^2/8$ for (e)-(f)). }\label{fig:1}
\end{figure*}

\begin{figure*}[htb]
\includegraphics[width=0.3\linewidth]{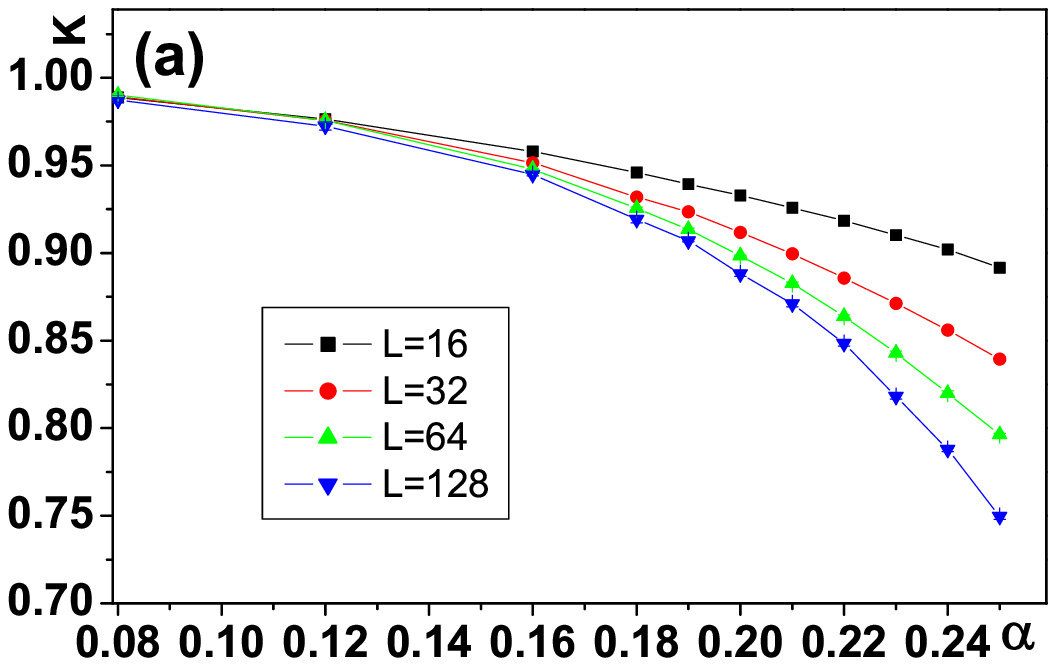}
\includegraphics[width=0.3\linewidth]{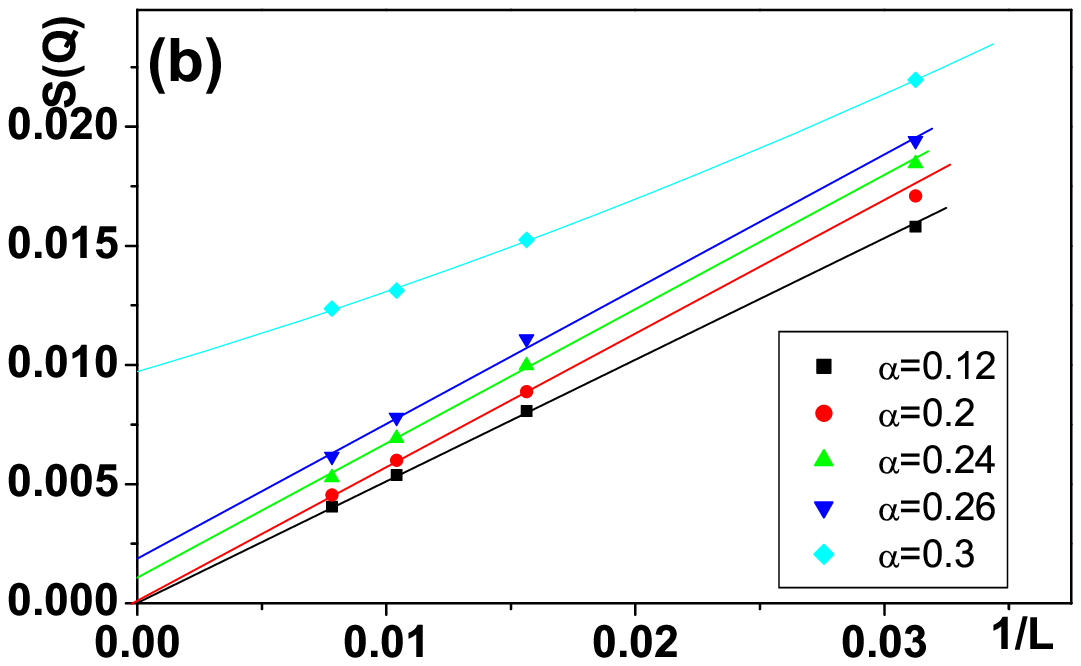}
\includegraphics[width=0.3\linewidth]{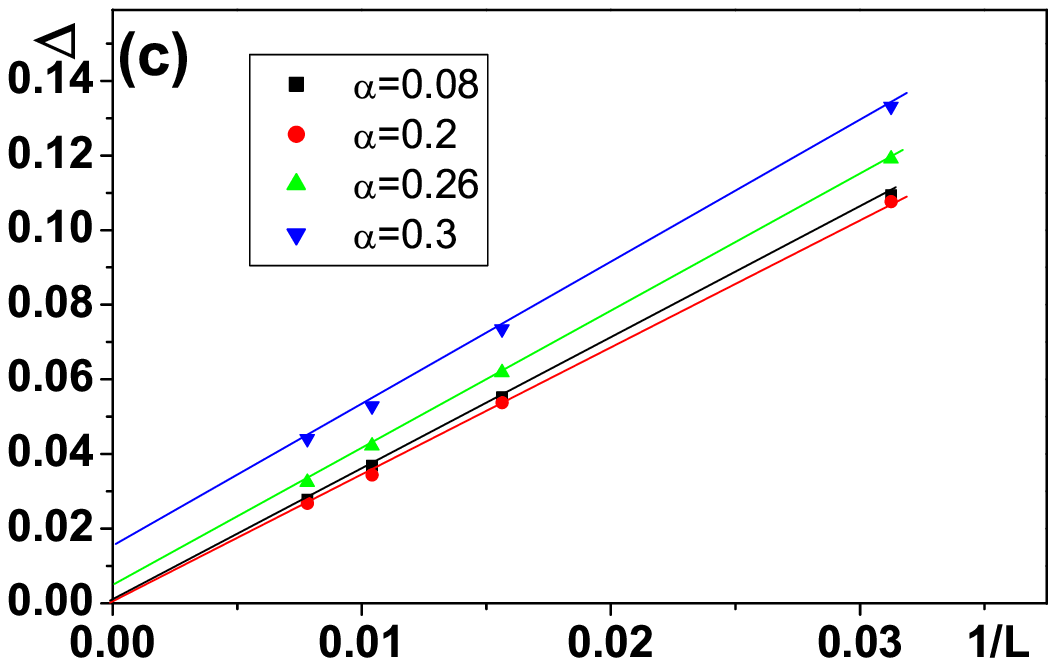}
\includegraphics[width=0.3\linewidth]{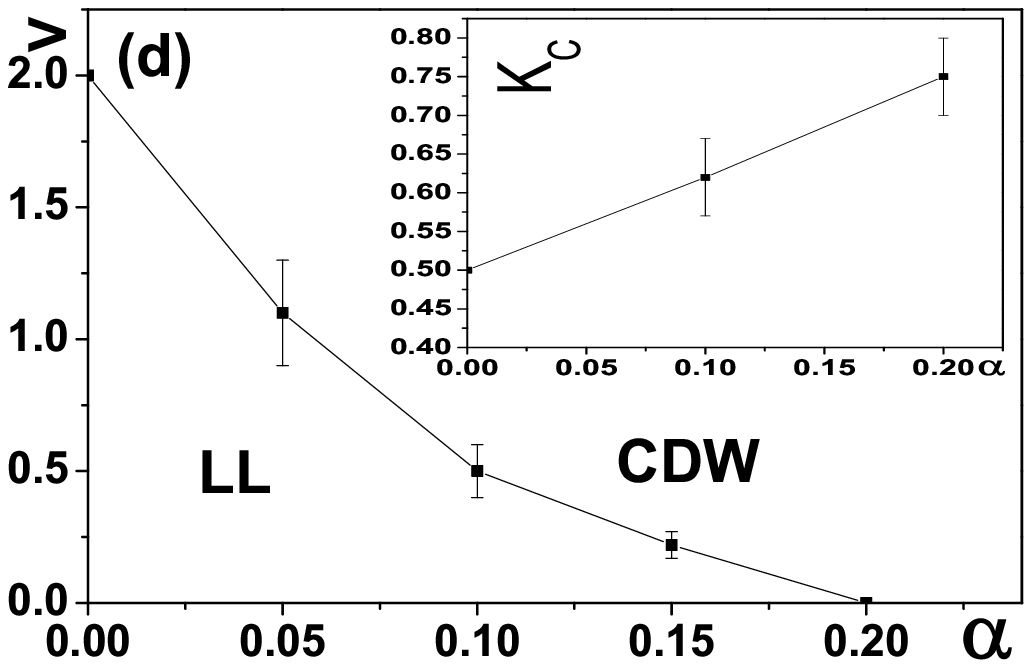}
\includegraphics[width=0.3\linewidth]{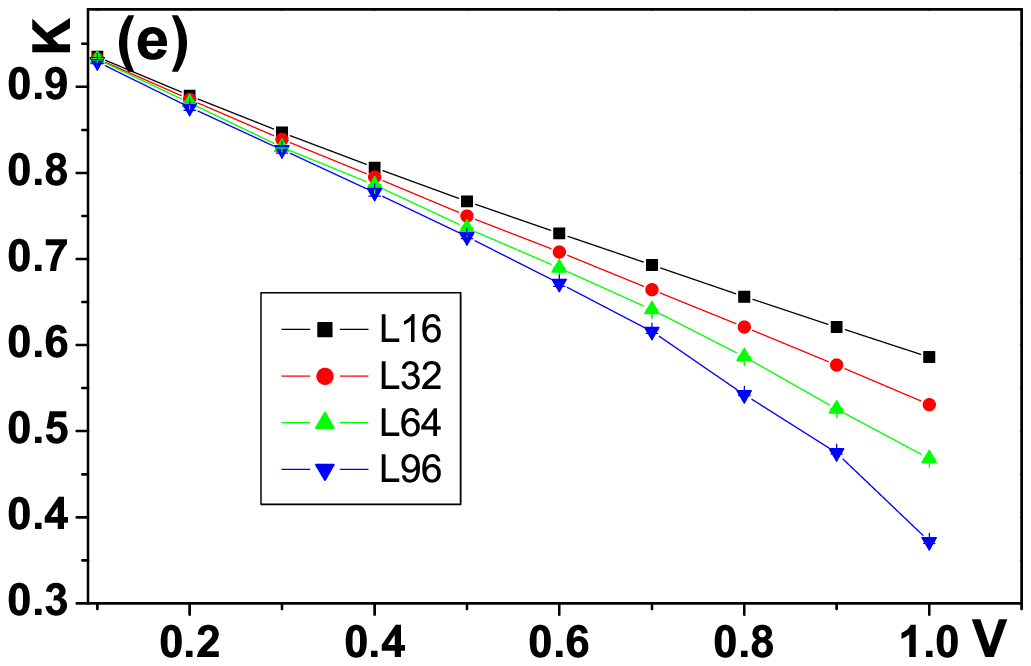}
\includegraphics[width=0.3\linewidth]{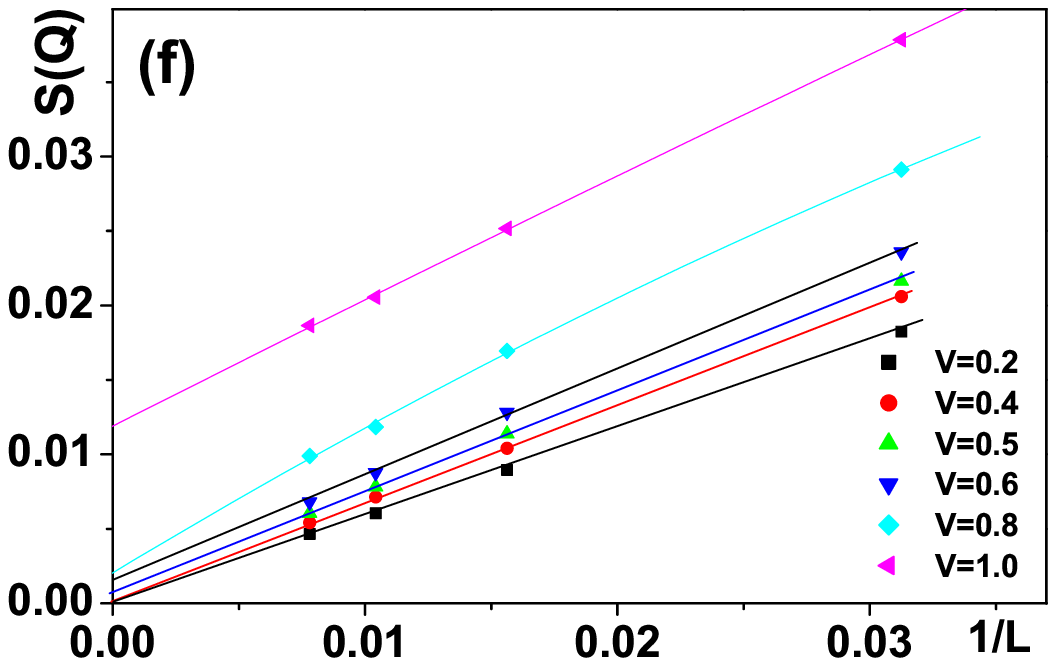}
\caption{(Color online). (a) Luttinger parameter $K$ as a function of $\alpha$ and
finite-size scaling of (b) the staggered structure factor $S(Q)$ and (c) the charge gap $\Delta$ for different $\alpha$ ($V=0$, $\mu=0$ for (a)-(c)). (d) Phase diagram at half-filling in the $\alpha-V$ plane (inset: the dependence of the critical Luttinger parameter $K_c$ on $\alpha$). (e) Luttinger
parameter $K$ as a function of $V$, and (f) finite-size scaling of
 $S(Q)$ for different $V$  ($\alpha=0.1$, $\mu=0$ for (e) and (f)).
Error bars are shown but may be smaller than the point size of the symbols; the scaling relation for (a)-(f) is $\beta=L/2$ for $z=1$.
} \label{fig:2}
\end{figure*}

{\it Incommensurate filling --}
We now switch to the discussion of the numerical results.
We first focus on the case of incommensurate filling of the hard-core bosons ($\mu\neq 0$). In the absence of dissipation, the physics is relatively straightforward: the groundstate is a Luttinger liquid (LL) irrespective of the interaction strength. To study the competition between quantum fluctuations and dissipation we set $V=0$ in Eq.(\ref{eq:hardcore}) and address the problem how the dissipation can qualitatively change the nature of the LL phase. To distinguish various quantum phases, we first study the single particle correlation function $G(r)=\langle a^\dag_i a_{i+r} \rangle$  for different $\alpha$. As shown in
Fig.\ref{fig:1} (a) and (b),  the single particle correlation function
decays algebraically for weak dissipation, while for strong dissipation it decays exponentially.
We also study the density correlation functions in (imaginary) time and space. We see that the on-site unequal-(imaginary) time  density correlation function $S(\tau)=\sum_i\langle (n_i(\tau)-\bar{n}) (n_i(0)-\bar{n})\rangle/L$ with $\bar{n}$ the density of the particle and $\bar{n}\approx 1/3$ (shown in Fig.\ref{fig:1} (c) and (d)) decays algebraically with $\tau$ for strong dissipation. This decay becomes however extremely weak with increasing $\alpha$; {\it e.g.,} for $\alpha=0.5, L=4$ and $\beta=108$ it is just $0.03$. Although this decay increases rapidly with $\beta$ we expect it to connect continuously to a constant in the limit $t/\alpha \to 0$.
The  equal-time density correlation functions  $S(r)=|\langle (n_i-\bar{n})(n_{i+r}-\bar{n})\rangle|$ decays algebraically with distance for weak dissipation, while for strong dissipation Fig.\ref{fig:1} (d) shows an exponential decay enveloping a density-dependent oscillatory factor, $S(r)\sim |e^{-\gamma r}\cos(2\pi r\bar{n})|$.
However, in contrast to $G(r)$, we find that $S(r)$ is much more sensitive to temperature. Based on a finite $\beta$ scaling of the factor $\gamma$  (see the inset of Fig.\ref{fig:2} d for $\alpha=0.36$), it is difficult to determine numerically whether $\gamma$ extrapolates to a very small but finite value or zero at zero temperature. Therefore, an algebraic decay ($\gamma \rightarrow 0$) for $S(r)$ at zero temperature is possible~\cite{Cazalilla2006}.
The different behaviors of the correlation functions for weak and strong dissipation clearly
indicate two distinct phases: in case of weak dissipation we have a Luttinger liquid (LL) while for strong dissipation we find the many-body counterpart of the localized phase in the spin-boson model.

To study the transition between the two distinct phases, we calculate other observables of interest such as the superfluid density $\rho_s$ and the compressibility $\kappa$ (or equilvalently the variance of winding number  $\langle W^2\rangle= \beta\rho_s/L $ ~\cite{Ceperley1987} and particle number $\Delta N=\langle N^2 \rangle- \langle N\rangle^2= L\kappa/\beta$, see the Suppl. Mat.~\cite{Supplementary}). In Fig.\ref{fig:1} (e), we plotted $\langle W^2 \rangle$ using the scaling relation with $z=2$, and found an intersection point between the different system sizes at the point $\alpha=0.33(1)$, indicating that the transition from LL to BIBL is not a Kosterlitz-Thouless type with $z=1$ as predicted by bosonization\cite{Cazalilla2006}, but a continuous transition with $z=2$ as in Ref.~\cite{Werner2005b}.
The $\rho_s$ is nonzero in the LL but approaches 0 in the BIBL phase. On the other hand, the variance of particle number, shown in Fig.~\ref{fig:1} (f), is larger in the BIBL than in the LL phase, indicating that the BIBL phase is a highly compressible phase with no charge gap. Furthermore, the BIBL has diffusive charge excitations resulting in a non-zero conductivity~\cite{Cazalilla2006} (also see the Suppl. Mat.~\cite{Supplementary}).

{\it Half filling --}
Now we turn to the half filled case ($\mu$=0) and focus on the XY model first ($V$=0) first. For weak dissipation, we find a LL phase just as in the incommensurate filling case. However, for strong dissipation, we find a Mott-insulator with CDW long-range order, which is characterized by an extensive staggered structure factor, found by Fourier transform of the density correlation function: $S(Q = \pi)=\frac
{1}{L^2}\sum_{i,j}(-1)^{i-j}\langle (n_i-\frac 12) (n_j-\frac
12)\rangle$. The finite-size scaling in Fig.\ref{fig:2}(b) indicates
that in the thermodynamic limit CDW long-range order emerges
for $\alpha>\alpha_c\approx 0.2$.
The dissipation-driven LL-to-CDW phase transition is reminiscent of a similar phase transition driven by the nearest-neighbor density-density interactions at constant density. Since the retardation is irrelevant in the LL phase, we analyze the transition from the LL side using LL terminology and anticipate a Kosterlitz-Thouless (KT) transition.

This can numerically be
verified from the dependence of the Luttinger parameter $K= \pi \sqrt{\rho_s \kappa}$ on $\alpha$,  as shown in Fig. \ref{fig:2} (a).  By performing a renormalization flow analysis (see
the Suppl. Mat.~\cite{Supplementary}), we can extract the position of the KT transition
point in the thermodynamical limit ($\alpha_c=0.20(1)$), which is
characterized by a sudden jump of the Luttinger parameter from
$K_c(L = \infty)=0.75(3)$ to $0$ determined via a Weber-Minnhagen fit~\cite{Weber1988}. The critical value of the Luttinger parameter $K_c$ cannot be understood from the lowest order renormalization-group equations~\cite{Malatsetxebarria2013} and a full explanation goes beyond the scope of this work. The KT phase transitions with non-universal $K_c$ have been observed in different contexts\cite{Horovitz2013,Pollet2014}.
Within our accuracy, the disappearance of $\rho_s$ coincides with the onset of the CDW order (and the charge gap (shown in Fig.\ref{fig:2} (c)).
However, the gap in the massive phase of the KT transition may be exponentially small and reading off such a gap is prone to error. Note that the CDW order is induced entirely by the dissipation, which reminds us of the Peierls transition in low-dimensional electron materials~\cite{Pouget2004}.

To complete the discussion and the phase diagram at half-filling, we also study the effect of NN interactions (as shown in Fig.\ref{fig:2} (d)). The situation
without dissipation ($\alpha=0$ in Eq.(\ref{eq:ohmic})) is well understood: the NN interactions can drive the
system from a LL to a CDW Mott-insulator at the
critical point $V_c=2t$ via a KT transition with $K_c = 1/2$. Turning on the dissipation suppresses quantum tunneling. We therefore expect that dissipation will make it easier for the system to access the CDW
Mott-insulating state. This is reflected in the numerics as is shown in Fig.\ref{fig:2} (c) and (d), where we see that for weak dissipation
($\alpha=0.1$) the phase transition point of the LL-CDW transition
is shifted down to $V_c=0.5t$. Along the phase boundary between the LL and the CDW, the critical Luttinger parameter changes continuously from $K_c = 0.5$ at $V=2$ (and $\alpha=0$) to $K_c = 0.75(3)$ at $V=0$ as is shown in the inset of Fig.~\ref{fig:2} (d). Within our accuracy, we saw no sign of an intermittent BIBL phase at half filling.

{\it Experimental realization and detection -- }
Hard-core bosons with Ohmic dissipation can be realized in a Bose-Fermi
mixture in an optical lattice by embedding quasi-1D  heavy bosons with strong repulsive interaction into a 3D fermi sea
composed of light fermions~\cite{Malatsetxebarria2013}.
The BIBL phase is characterized by the exponential decay of the single-particle correlation function with distance, which can be
seen in time-of-flight interference experiments. The finite
compressibility and the density-density correlation function can be measured with in-situ single-site resolution techniques~\cite{Bakr2010,Sherson2010}.
Conductivity measurements would require phase modulation of the lattice\cite{Tokuno2011}.

{\it Conclusion and outlook -- }
In summary, we generalize the worm algorithm to study a 1D open quantum many-body model consisting of hard-core bosons where the density of every particle couples Ohmically to an independent, local bath. Away from half filling, we found a homogeneous, compressible, conducting but non-superfluid bath-induced Bose liquid phase, which can be seen as the many-body generalization of the localized states in the spin-boson model. At half-filling, we find a KT type phase transition between the CDW and LL phases, but with a critical value of the Luttinger parameter that is in general non-universal.  Our method can be applied to all open bosonic and unfrustrated spin systems with a similar form of the density-type coupling to the bath, in one or higher dimensions, and with Ohmic or non-Ohmic dissipation.  In future work, the generalization of our method to higher
dimensional systems, or systems with a different Hamiltonian (e.g. gapped systems) or different type of dissipation (e.g. sub-ohmic) will be studied, as well as the entanglement properties.

{\it Acknowledgements -- } We wish to thank  M. Cazalilla, I. Cirac, T. Giamarchi, and B. Svistunov for fruitful discussions. This work was supported in part by the German Research Foundation under DFG FOR801, by FP7/Marie-Curie Grant No. 321918, FP7/ERC Starting Grant No. 306897, and in part by the National Science Foundation under Grant No. PHYS-1066293 and the hospitality of the Aspen Center for Physics and the Kavli Institute for Theoretical Physics China. Z. C. also acknowledges the support from
Austrian Science Fund through SFB FOQUS (FWF Project
No. F4006-N16) and the ERC Synergy Grant UQUAM.


\end{document}